\documentclass[a4paper]{article}

\usepackage[utf8]{inputenc}
\usepackage{amsmath}
\usepackage{amssymb}
\usepackage{bm}
\usepackage{units}

\newcommand{\lP}{\ensuremath{\ell_P}}
\newcommand{\lB}{\ensuremath{\ell_B}}

\title{Violation of the Weak Equivalence Principle in Bekenstein's
  theory} 
\author{L. Kraiselburd,  H. Vucetich\\
\emph{FCAGLP--UNLP, Argentina}}

\begin{document}
\maketitle
\bibliographystyle{unsrt}

\begin{abstract}
  Bekenstein has shown that violation of Weak Equivalence Principle is
  strongly supressed in his model of charge variation. In this paper,
  it is shown that nuclear magnetic energy is large enough to produce
  observable effects in E\"otv\"os experiments.
\end{abstract}

\section{Introduction}
\label{sec:Intro}

The variation of fundamental constants has been an important subject
of research since Dirac stated the Large Number Hypothesis (LNH)
\cite{Dirac37,Dirac38}. In the latter times, the interest in that
subject has been aroused again since such a variation is a common
prediction of several ``Theories of Everything'' (TOEs), such as
string theories \cite{Uzan02}. One possible low-energy limit of these
TOEs is Bekenstein's variable charge model
\cite{Beckens82,Bekenstein:2002wz,SBM02}, since it has all desirable
properties that such low-energy limit should exhibit.

Since Dirac's proposal, many attempts have been made to detect the
proposed variations, most of them with null results (for reviews, see
\cite{Uzan02,TVOFC1,Vucetich:2002uq}). An interesting possibility is
that a space variation of fundamental constants should produce a
violation of the Weak Equivalence Principle \cite{MThW}, a fact that
can be proved easily using energy conservation \cite{Haugan79}.

The Weak Equivalence Principle (WEP) states that the world line of a
body immersed in a gravitational field is independent of of its
composition and structure \cite{MThW,WillTh}, a generalisation of
Galileo's law of Universality of Free Fall: the local acceleration
$\bm{g}$ of a body is independent of its composition and
structure. Since General Relativity has the Equivalence Principle as
one of its consequences, testing for its validity is an important form
of the search of ``new physics''.

The most sensitive forms of those tests are the \emph{E\"otv\"os
  experiments}: testing the equality of acceleration for bodies of
different of different composition or structure
\cite{EPF22,Weinberg,MThW}.  Several accurate tests have been carried
in the second half of the 20th century and up to now
\cite{RKD64,Braginski72,KeiFall82,Su94,%
  Baessler:1999zz,Schlamminger:2007ht}.
These tests impose strict bounds on parameters describing WEP
violations \cite{WillTh,Chamoun:1999pe}.

However, in his 2002 paper Bekenstein \cite{Bekenstein:2002wz} proved
that a violation of WEP is highly unlikely in his model. We shall
discuss briefly this issue later on, but the origin of this statement
is a wonderful cancellation of electrostatic sources of the $\psi$ field,
leading to a null effect in the lowest order. No such cancellation
happens for magnetostatic contribution, but a simple examination of
the Solar System magnetic energy density suggests that a breakdown of
WEP should be inobservable.

In this paper, we discuss the detection of a space variation of $\alpha$ in
Bekenstein's model, considering the fluctuations of magnetic fields in
quantum systems. 

The rest of the paper is structured as follows: in section
\ref{sec:Bek:Model} we make a short summary of the main results on
Bekenstein's model related to our problem; section
\ref{sec:Bek:WEP:Th} deals with the motion of a composite
nonrelativistic body in external gravitational plus dilaton fields, to
find an expression for its anomalous acceleration; in section
\ref{sec:Bek:MagE} we discuss the magnetostatic energy of matter in a
quantum system and in section \ref{sec:Results} we state our results
and conclusions. The appendix is devoted to a simple proof of equation
(\ref{eq:Bek:Mag:Ener}).

\section{A survey of Bekenstein model}
\label{sec:Bek:Model}

Bekenstein's proposal \cite{Beckens82,Bekenstein:2002wz} was to modify
Maxwell's electromagnetic theory introducing a field $\epsilon$ to describe
$\alpha$ variation. An unique form of the theory (up to a parameter) was
found from the following hypotheses:
\begin{enumerate}
\item The theory should reduce to Maxwell's for a constant $\alpha$.
\item $\alpha$ variation must be dynamical (i.e. generated by a field $\psi =
  \ln \epsilon $).
\item The dynamics of the field is derived from a variational
  principle.
\item The theory must be causal, gauge and time-reversal invariant.
\item \label{item:lP} The smallest length in the theory should be the
  Planck length \lP.
\end{enumerate}

The latter statement should be dropped if the theory is considered a
low-energy limit of some TOE, since these introduce other fundamental
length scales.

The application of the above hypothesis lead to a unique form of the action
\begin{equation}
  \label{eq:Bek:Stot}
  S = S_{\rm em} + S_{\psi} + S_{\rm mat} + S_G,
\end{equation}
where
\begin{subequations}
  \begin{equation}
S_{\rm em} = -\frac{1}{16\pi}\int e^{-2\psi}f^{\mu\nu}f_{\mu\nu} \sqrt{-g} d^4x 
    \label{eq:Bek:Max}
    \end{equation}
    is the modified Maxwell action $\left[f_{\mu\nu} = (e^{\psi}A_{\nu})_{,\mu} -
      (e^{\psi}A_{\mu})_{,\nu} = e^{\psi}F_{\mu\nu}\right]$; 
    \begin{equation}
    S_{\psi} = \frac{-\hbar c} {2\lB^2}  \int ( \partial_\mu\psi )^2 \sqrt{-g} d^4x 
    \label{eq:Bek:psi}      
    \end{equation}
  \label{eq:Bek:S_i}
\end{subequations}
is the $\psi$ field action and $S_{\rm mat}$ and $ S_G $ are the matter
and Einstein actions. The local value of the electric charge is
\begin{align}
  \label{eq:local:q}
  e(x^\mu) &= e_0 e^{\psi(x^\mu)} & \alpha(x^\mu) &= e^{2{\psi(x^\mu)}} \alpha_0
\end{align}
where $e_0, \alpha_0$ are reference values of these magnitudes. 

The general equations of motions for these fields are
\begin{subequations}\label{eq:Beq:em}
  \begin{gather}\label{eq:Beq:Max:em}
    \left(e^{-\psi}F^{\mu\nu}\right)_{,\nu} = 4\pi j^\mu, \\
    \Box \psi = \frac{\lB^2}{\hbar c} \left(\frac{\partial\sigma}{\partial\psi}
      -\frac{F^{\mu\nu}F_{\mu\nu}}{8\pi} \right), \label{eq:Beq:psi:em}
    \\
    \sigma=\sum mc^{2}\gamma^{-1}(-g)^{-1/2}\delta[\bm{x} -
    \bm{x}(\tau)], \label{eq:Beq:def:sigma} 
  \end{gather}
\end{subequations}
the latter quantity being the rest mass energy density.

A word of advice is due: in his papers
\cite{Beckens82,Bekenstein:2002wz} Bekenstein uses an ensemble of
classical particles to represent matter. This is not a good model of
matter wherever quantum phenomena are important, neither at high
energy scales or small distances scales, since fermions have a
``natural length scale'', namely the Compton wavelength of the
particle $\lambda_C = \hbar/mc$. One must be wary of jumping to
conclusions in these regimes. (See also \cite{Damour03:vc}).

From the above equations of motion Bekenstein \cite{Bekenstein:2002wz}
derives several theorems.
\begin{description}
\item[Cancellation theorem] For a electrostatic field equation
  (\ref{eq:Beq:psi:em}) can be written in the form
  \begin{subequations}\label{eq:Bek:em:stat}
    \begin{gather}
      \label{eq:Bek:em:Est}
      \bm{\nabla\cdot}(e^{-2\psi} \bm{E}) = 4\pi\rho\\
      \nabla^2\psi = 4\pi\kappa^2 \left[\frac{\partial\sigma}{\partial\psi} + e^{-2\psi}\frac{E^2}{4\pi},
      \right] \label{eq:Bek:em:psi:Est}
      \\ 
      \kappa^2 = \frac{\lB^2}{4\pi\hbar c}. \label{eq:Bek:def:kappa}
    \end{gather}
  \end{subequations}
  
  In the source term for $\psi$ the first term cancels almost exactly
  the second and the asymptotic value of $\psi$ is almost exactly
  suppressed.
\item[WEP for electric charges] The equation of motion of a system of
  charges in an electric field, in the limit of very small velocities,
  reduces to
  \begin{equation}
    \label{eq:Bek:em:charges}
    M \ddot{\bm{Z}} = Q \bm{E}
  \end{equation}
  and where $M$ and $Q$ are the total mass and charge of the
  system. Thus, there is no WEP violation.
\end{description}

The above results use the classical point charges model of matter. On
the other hand, the equation of motion for $\psi$ for a static system of
magnetic dipoles is
\begin{equation}\label{eq:Bek:em:psi:Bst}
  \nabla^2\psi = -4\pi\kappa^2e^{-2\psi}\frac{B^2}{4\pi}
\end{equation}
and there is no cancellation of sources. From an estimate of the field
intensities in the Solar System, Bekenstein states that no observable
WEP violation can be detected in laboratory experiments. This latter
result is also based in the classical point charges model of matter.

\section{Motion of a composite body in the $\psi$ field}
\label{sec:Bek:WEP:Th}

Let us now find the Lagrangian of a body composed of point-like
charges, such as an atom or an atomic nucleus. We shall work in the
nonrelativistic limit for the charges, but we shall keep for the
moment the full expression for the electromagnetic field. We shall
treat the system as classical and later on quantize it in a simple
way. The techniques we use are a lightweight version of those used in
the $TH\epsilon\mu$ formalism \cite{LightLee73,WillTh}.
We assume that there are external dilaton $\psi$ and Newtonian
gravitationa $\phi_N$ fields acting over
the body, but we shall neglect the self fields generated. With these
approximations, the Lagrangian of the system takes the form
\begin{equation}
  \label{eq:Beq:TestBody:L}
  \begin{split}
    L =& - M_{\rm tot}\left[\psi\right]c^2\\
    &+ \sum_a \left[ \frac{1}{2} m_a v_a^2 - m_a\phi_N(\bm{x}_a) - e_a\Phi(\bm{x}_a)
      - \frac{e_a}{c} \bm{v}_a \bm{\cdot A}(\bm{x}_a)\right]\\
      &- \int \frac{e^{-2\psi}(E^2 - B^2)dV}{16\pi}.
    \end{split}
\end{equation}

To eliminate the electromagnetic fields we use the equations of motion
(\ref{eq:Beq:Max:em}) together with the Lorentz gauge condition to
obtain
\begin{align}
  \label{eq:Bek:em:Amu}
  F^{\mu\nu}_{,\nu}&= \frac{4\pi}{c} j^\mu_{\rm ef}, & j^\mu_{\rm ef} &= e^{\psi}
  \left(j^\mu  + \frac{c}{2\pi} \psi_{,\nu}F^{\mu\nu}\right).  
\end{align}

These equations can be solved using retarded potentials
\begin{align}
  \label{eq:Bek:RetPot}
  \Phi(\bm{x},t) &= \int \frac{\rho_{\rm ef}(t_{\rm ret})}{R} dV', &
  \bm{A}(\bm{x},t) &= \frac{1}{c} \int \frac{\bm{j}_{\rm ef}(t_{\rm
      ret})}{R}dV', & R = \mid \bm{x} - \bm{x}'\mid, 
\end{align}
whose slow-motion approximations are
\begin{align}
  \label{eq:Bek:Pot:Slow}
  \Phi &= \int \frac{\rho_{\rm ef}}{R} dV', & \bm{A} &= \frac{1}{c} \int
  \frac{\bm{j}_{\rm ef}}{R}dV', & \bm{B} &= \frac{1}{c} \int
  \frac{\bm{j}_{\rm ef} \bm{\times R} }{R^3}dV.
\end{align}

In these equations we shall neglect the contribution of $\psi_{,\nu}$ since
they are much smaller than the usual current contribution. After some
transformations the Lagrangian can be written as
\begin{equation}
  \label{eq:Bek:Lag:Body:Fin}
  \begin{split}
    L =& - M_{\rm tot}\left[\psi\right]c^2\\
    &+ \sum_a \left[ \frac{1}{2} m_a v_a^2 - m_a\phi_N(\bm{x}_a)\right] 
      - \frac{1}{2}\int e^{2\psi}\frac{\rho_c(\bm{x})\rho_c(\bm{x}')}{R}
      dV dV'\\
      &+ \frac{1}{2c^2}\int e^{2\psi}\frac{\bm{j}(\bm{x})\bm{\cdot
          j}(\bm{x}')}{R} dV dV' .
  \end{split}
\end{equation}
where we have replaced sums over pair of particles or currents with
integrals. 

For a macroscopic solid body we shall be interested in the motion of
the center of mass. The separation of this motion is easily achieved
with the usual substitutions and developing the slowly varying
external fields $\phi_N$ and $\psi$:
\begin{align}
  \label{eq:Bek:CM:Body}
  \bm{R}_{\rm CM} &= \frac{\sum_a m_a\bm{x}_a} {M_{\rm tot}}, &
  \bm{V}_{\rm CM} &= {\sum_a \bm{v}_a},\\
  \bm{x}_a &= \bm{R}_{\rm CM} + \bm{x}'_a, & \bm{v}_a &= \bm{V}_{\rm
    CM} + \bm{v}'_a,\\
  \phi_N(\bm{x}) &\simeq \phi_N(\bm{R}_{\rm CM}) + O\left({x'}^2\right), &
  \psi(\bm{x}) &\simeq \psi(\bm{R}_{\rm CM}) + O\left({x'}^2\right);
\end{align}
and besides
\begin{equation}
  \label{eq:M-exp}
  M_{\rm tot}\left[\psi\right]c^2 \simeq M_{\rm tot}\left[0\right] +
  \frac{\partial Mc^2}{\partial\psi} \psi(\bm{R}_{\rm CM}) + O\left({\psi}^2\right).
\end{equation}

Substitution of the above in the Lagrangian leads to
\begin{equation}
  \label{eq:Bek:Lag:CM:Body}
  \begin{split}
    L =& -{M_{\rm tot}}\left[c^2 - \frac{V_{\rm CM}^2}{2} -
      \phi_N(\bm{R}_{\rm CM}) +
      \right] \\
    &+ 2 \psi(\bm{R}_{\rm CM}) {E_m} + \dots
  \end{split}
\end{equation}

The electrostatic contribution cancels with the mass dependence on
$\psi$, according to Bekenstein theorem, and the neglected terms are
of either tidal order, negligible in laboratory tests of WEP, or of
higher order in $\psi$.

The above Lagrangian shows that a body immersed in external
gravitational and Bekenstein fields will suffer an acceleration
\begin{equation}
  \label{eq:Bek:Acc:Body}
  \ddot{\bm{R}}_{\rm CM} = \bm{a} = \bm{g} + 2 \frac{{E}_m}{M}
  \left.\bm{\nabla}\psi \right\vert_{\rm CM}. 
\end{equation}

The latter term is the anomalous acceleration generated by the
Bekenstein field.  The acceleration difference (\ref{eq:Bek:Acc:Body})
is tested in E\"otv\"os experiments.

\section{Magnetic energy of matter}
\label{sec:Bek:MagE}

In a quantum model of matter, magnetic fields originate in not only in
the stationary electric currents that charged particle originate and
their static magnetic moments but also in quantum fluctuations of the
number density. These contributions to the magnetic energy have been
computed in Ref. \cite{Haugan:1977px,WillTh} from a minimal nuclear
shell model. The matrix elements of the current operator can be
related to the strength of the dipole resonance, with the result
\begin{equation}
  \label{eq:Bek:Mag:Ener}
  E_m = \int d^3x \frac{B^2}{8\pi}  \simeq \frac{1}{2c^{2}} \int d^3x d^3x'
  \frac{\bm{j}(\bm{x})\cdot\bm{j}(\bm{x}')}{\mid \bm{x} - \bm{x}'\mid } \simeq
  \frac{3}{20\pi} \frac{\hat{E}}{R(A)\hbar c} \int\sigma dE,
\end{equation}
where $R(A)$ is the nuclear radius, $\hat{E}$ is the giant dipole
mean absorption energy and $\int\sigma dE$ its integrated strength
function. These quantities have the following approximate
representation
\begin{align}
  \label{eq:Bek:Mag:E:Par}
  R(A) &=  \unit[1.2 A^{\frac{1}{3}}]{fm}, & \hat{E} &\sim
  \unit[25]{MeV}, 
  &  \int\sigma dE &\simeq  \unit[1.6 A]{MeV\; fm^2}.
\end{align}

Since  the magnetic energy density  is concentrated near atomic
nuclei, it can be represented in the form
\begin{equation}
  \label{eq:Bek:Mag:Eden}
  e_m(\bm{x}) = \sum_a E_m^a \delta(\bm{x} - \bm{x}_a) \simeq
  \sum_b{E}_m^b n_b(\bm{x}), 
\end{equation}
where index $b$ runs over different nuclear species. Define
\begin{equation}
  \label{eq:Bek:Def:zeta}
  \zeta_m^b = \frac{E_m^b}{M_bc^2}
\end{equation}
as the fractional contribution of the magnetic energy to rest mass. Then
\begin{equation}
  \label{eq:Bek:Mag:Eden2}
  e_m(\bm{x}) = \bar{\zeta}_m(\bm{x}) \rho(\bm{x})c^2,
\end{equation}
where $\rho(\bm{x})$ is the local mass density and
\begin{equation}
  \label{eq:Bek:Def:zetabar}
  \bar{\zeta}_m(\bm{x}) = \frac{\sum_b \zeta_b \rho_b(\bm{x})}{\rho(\bm{x})}
\end{equation}
is the local mass-wheighted average of $\zeta_m$.

With expression (\ref{eq:Bek:Mag:Eden2}) we can write equation
(\ref{eq:Bek:em:psi:Bst}) in the form
\begin{equation}
  \label{eq:Bek:em:psi:rho}
  \nabla^2\psi = -8\pi\kappa^2c^2 e^{-2\psi}\bar{\zeta}_m \rho.
\end{equation}
For small $\psi$ we can find a solution for an arbitrary distribution of
sources
\begin{equation}
  \label{eq:Bek:Sol:psi}
    \psi 
    = 8\pi\kappa^2c^2 \frac{1}{r} \int_0^r x^2 \bar{\zeta}_m(x) \rho(x) dx
\end{equation}
whose asymptotic behaviour can be expressed in terms of the newtonian
gravitational potential
\begin{equation}
  \label{eq:Bek:Sol:psi:fi}
  \psi \asymp  \frac{8\pi\kappa^2}{GM} \phi_N(r) \int_0^\infty x^2
  \bar{\zeta}_m(x)   \rho(x) dx = 
   2 \left(\frac{\lB}{\lP} \right)^2 \tilde{\zeta}_m
  \frac{\phi_N(r)}{c^2}, 
\end{equation}
where $\tilde{\zeta}_m$ is the mass averaged value of $\zeta_m$ and we
have introduced the Planck length \lP. 

\section{Results and conclusion}
\label{sec:Results}

 From \eqref{eq:Bek:Sol:psi:fi} we obtain for the
differential acceleration of a pair $A, B$ of different bodies
\begin{equation}
  \label{eq:Bek:eta:expr}
  \eta(A, B) = \frac{a_A - a_B}{g} =4 \left(\frac{\lB}{\lP}
  \right)^2 \zeta_S (\zeta_{A} - \zeta_{B}) = C_f \left(\frac{\lB}{\lP}
  \right)^2
\end{equation}
where $\zeta_S, \zeta_{A},  \zeta_{B} $ are the magnetic energy fractions
of the source, body $A$ and body $B$ respectively.
\begin{equation}
  \label{eq:Bek:Em:Def}
  \zeta_I = \frac{E_I}{M_Ic^2}
\end{equation}

Table \ref{tab:EotRes} shows the results of the most accurate versions
of the E\"otv\"os experiment. A simple least squares fit with the
statistical model $y = C_f x^2$ yields
\begin{equation}
  \label{eq:Res1:lBlP2}
  \left(\frac{\lB}{\lP}\right)^2 = 0.0003 \pm 0.0006 
\end{equation}
from which we get the ``$3\sigma$'' upper bound
\begin{align}
  \label{eq:Res2:UppBnd}
   \left(\frac{\lB}{\lP}\right)^2 <& 0.002 & 
  \frac{\lB}{\lP} &< 0.05 
\end{align}

This last equation encodes the main result of this paper: strict upper
bounds can be set from E\"otv\"os experiments on the Bekenstein parameter
$\lB/\lP$ even if the electrostatic field does not generate $\psi$
field. These bounds are much larger than the ones that would result if
electrostatic energy density would generate $\psi$ field intensity. This
calculation was carried in the 1982 paper of Bekenstein
\cite{Beckens82} and has been repeated several times
(e.g. \cite{Dvali:2001dd,Chamoun:1999pe,mosquera08:tvbek}) with the
result
\begin{equation}
  \label{eq:Res:Old}
  \left(\frac{\lB}{\lP}\right)_{\rm el} < 8.7\times 10^{-3},
\end{equation}
one order of magnitude smaller than (\ref{eq:Res2:UppBnd}).

It is interesting to compare our result (\ref{eq:Res2:UppBnd}) with
the results obtained from an analysis of all evidence from time
variation of the fine structure constant $\alpha$
\cite{mosquera08:tvbek}. In that paper, an effective value of
$\zeta=10^{-4}$ was used, following the suggestion of ref. \cite{SBM02}
and a $1\sigma$ bound on $(\lB/\lP)^2 < 0.003$ was found.  From the
estimate of $\zeta_{\rm H}$ in reference \cite{Bekenstein:2002wz} we
compute an effective value of $\zeta_U = 2.7\times10^{-5}\Omega_B \simeq 1.4\times10^{-6}$ and
so we find a $3\sigma$ upper bound
\begin{equation}
  \label{eq:Bek:Upp:TV}
  \frac{\lB}{\lP} < 0.8
\end{equation}
one order of magnitude larger than \eqref{eq:Res2:UppBnd}.

In conclusion, we have shown that very strict bound can be put on the
Bekenstein model parameter $\lB/\lP$ from the quantum fluctuations of
the magnetic fields of matter. 
From equation (\ref{eq:Res2:UppBnd}) one should discard the Bekenstein
model, but since it can be obtained as a low energy limit of string
models, the latter conclusion should be taken with a grain of salt.

\section*{Acknowledgements}

We are grateful to Drs.S.J. Landau and P.D. Sisterna  for their
interesting suggestions and comments. L. K. is fellow of CONICET.

 \begin{table}
   \centering
   \begin{tabular}{ccccr@{$\pm$}lr}
     \hline
     $A$ & $B$ & Source & $10^{11}C_f$ &
     \multicolumn{2}{c}{$10^{11}\eta(A,B)$} & Ref.\\  
     \hline
     Al & Au & Sun & 17.5 & 1.0 & 1.5 & \cite{RKD64}\\
     Al & Pt & Sun & 17.5 & 0.03 & 0.045 & \cite{Braginski72}\\
     Cu & W  & Sun & 8.8 & 0.0 & 2.0 & \cite{KeiFall82}\\
     Be & Al & Earth & 6.8 & -0.02 & 0.23 & \cite{Su94}\\
     Be & Cu & Earth & 10.4 & -0.19 & 0.25 & \cite{Su94}\\
     Be & Al & Sun & 16.1 & 0.40 & 0.98 & \cite{Su94}\\
     Be & Cu & Sun & 24.6 & -0.51 & 0.61 & \cite{Su94}\\
     Si/Al & Cu & Sun & 8.8 & 0.51 & 0.67  & \cite{Su94}\\
     EC & MM &  Sun & -7.6 & 0.001 & 0.032 & \cite{Baessler:1999zz}\\
     Be & Ti & Earth & 6.9 & 0.004& 0.018 & \cite{Schlamminger:2007ht}\\
     \hline
   \end{tabular}
   \caption{Results of E\"otv\"os experiments. The columns show the
     composition of the bodies, the source, the coefficient of
     $(\lB/\lP)^2$ in equation (\ref{eq:Bek:eta:expr}), the measured
     value of $\eta$ and its $1\sigma$ error.}
   \label{tab:EotRes}
 \end{table}

\appendix
\section{Proof of (\ref{eq:Bek:Mag:Ener})}

Reference \cite{Haugan:1977px} does not give a proof of equation
(\ref{eq:Bek:Mag:Ener}). The following proof is based on their
methods.

Let us write the total magnetic energy of the nucleus in the form
\begin{equation}
  \label{eq:Mag:Ener:Expr}
  E_m \frac{1}{2c^2}\sum_\alpha \int d\bm{x} d\bm{x}' \frac{\left\langle0\right\vert
    \bm{j}(\bm{x})\left\vert \alpha\right\rangle \bm{\cdot}
    \left\langle\alpha\right\vert \bm{j}(\bm{x}) \left\vert0\right\rangle }%
  {\vert \bm{x} - \bm{x}' \vert},
\end{equation}
where $\alpha$ runs over a complete set of eigenstates of the nuclear
hamiltonian $H$. The current operator is defined as
\begin{equation}
  \label{eq:Bek:Curr:Op:Def}
  \bm{j}(\bm{x}) = \sum_a \delta\left(\bm{x} - \bm{x}_a\right)
  e_a\frac{\bm{p}_a}{m_a}, 
\end{equation}
where the sum runs over all particles in the system. Neglecting the
momentum dependence of the nuclear potential, we can write
\begin{equation*}
  \frac{\bm{p}_a}{m_a} = \frac{i}{\hbar}\left[x_a,H\right].
\end{equation*}

Substitution of the above in equation \eqref{eq:Bek:Curr:Op:Def} yelds
the result
\begin{equation}
  \label{eq:Bek:Curr:Dipole}
  \begin{split}
      \left\langle0\right\vert \bm{j}(\bm{x}) \left\vert\alpha\right\rangle &=
  \frac{i}{\hbar} \sum_a \delta\left(\bm{x} - \bm{x}_a\right) (E_0 - E_\alpha)
  \left\langle0\right\vert e_a\bm{x}_a\left\vert\alpha\right\rangle \\
  &= \frac{i}{\hbar} \sum_a \delta\left(\bm{x} - \bm{x}_a\right) (E_0 -
  E_\alpha) \bm{d}_{0\alpha},
  \end{split}
\end{equation}
with $\bm{d}(\bm{x})$ the polarization (dipole density) operator. 

If we assume a constant density within the nucleus, the dipole density
can be represented as
\begin{equation*}
  \bm{d}_{0\alpha} = \frac{d_{0\alpha}}{V_N} \hat{\bm{x}}
\end{equation*}
where $V_N = \frac{4\pi}{3} R_N^3 $ is the nuclear volume, and so
\begin{equation}
  \label{eq:Bek:Curr:Prod}
  \left\langle0\right\vert \bm{j}(\bm{x}) \left\vert\alpha\right\rangle \bm{\cdot}
  \left\langle\alpha\right\vert \bm{j}(\bm{x}) \left\vert0\right\rangle \simeq
  \frac{\left\vert d_{0\alpha} \right\vert^2}{\hbar^2}
  \frac{E_{0\alpha}^2}{V_N^2} \cos\theta, 
\end{equation}
where $\theta$ is the angle between $\hat{\bm{x}}$ and
$\hat{\bm{x}}'$. Thus, the magnetic energy can be expressed
approximately as
\begin{equation}
  \label{eq:Mag:Ener:Expr:2}
  E_m \simeq \frac{\sum_a E_{0\alpha}^2 \left\vert d_{0\alpha}\right\vert^2}{2\hbar^2c^2}
        \frac{\int d\bm{x}d\bm{x}' \frac{\cos\theta}{\vert \bm{x} -
            \bm{x}' \vert}}{V_N^2}.
\end{equation}

The last factor is equal to $\frac{3}{5R_N}$. The first one can be
computed from the connection between the strength function and the
photoabsorption cross section
\begin{equation}
  \label{eq:Def:Strength:Function}
  \sigma_{0\alpha} = \frac{4\pi}{\hbar c} E_{\alpha0} \vert d_{\alpha0} \vert^2.
\end{equation}

From this, we easily get
\begin{equation}
  \label{eq:Mom:Strength}
  \sum_a E_{\alpha0}^2 \vert d_{\alpha0} \vert^2 = \frac{\hbar c}{4\pi}
  \frac{\int E\sigma(E)dE}{\int\sigma(E)dE} \cdot  \int\sigma(E)dE = \bar{E} \int\sigma(E)dE
\end{equation}
where $\bar{E} \sim \unit[25]{MeV}$ is the mean absorption energy,
roughly independent of $A$. 

The cross section satisfies the Thomas-Reiche-Kuhn 
sum rule
\begin{equation}
  \label{eq:TRK:Sum:Rule}
  \int\sigma(E)dE = (1+x)\frac{2\pi^2e^2\hbar}{mc} \frac{NZ}{A} \simeq
  (1+x)\unit[15]{MeV\,mbarn} A,
\end{equation}
where $x \sim 0.2$ takes into account exchange and velociy dependence of
nuclear interactions. Combining equations \eqref{eq:Mag:Ener:Expr:2},
\eqref{eq:Mom:Strength} and \eqref{eq:TRK:Sum:Rule} we obtain
equations \eqref{eq:Bek:Mag:Ener} and \eqref{eq:Bek:Mag:E:Par}.

\bibliography{tvref_ph,tvref_th,tvref_var,gener,Eotvos}
\end{document}